\documentclass[]{spie}  

\usepackage{amsmath,amsfonts,amssymb}
\usepackage{graphicx}
\usepackage[colorlinks=true, allcolors=blue]{hyperref}
\usepackage{soul}
\usepackage{upgreek}
\usepackage{lineno}

\newcommand{\um}{$\upmu$m}
\newcommand{\us}{$\upmu$s}
\newcommand{\arcsec}{$^{\prime\prime}$}

\title{96 kHz on-sky imaging on an adaptive optics system with a single-photon avalanche diode}

\author[*a]{Theodore A. Grosson}
\author[b]{Maaike A. M. van Kooten}
\author[b,a]{Kathryn Jackson}
\author[b]{Jean-Pierre Véran}
\author[c]{Simon Carrier}
\affil[a]{University of Victoria, 3800 Finnerty Rd, Victoria, Canada}
\affil[b]{NRC Herzberg Astronomy and Astrophysics Research Centre, 5071 W Saanich Rd, Victoria, Canada}
\affil[c]{Université de Sherbrooke, 2500 Boul de l'Université, Sherbrooke, Canada}

\authorinfo{*tgrosson@uvic.ca}

\begin{document} 
\maketitle


\begin{abstract}
Astronomical observations requiring extremely high angular resolution necessitate advanced adaptive optics (AO) systems to overcome blurring caused by the atmosphere. In addition to obtaining much sharper point-spread functions (PSFs) with these systems, it is beneficial to be able to characterize the behaviour of the resulting PSF across time and wavelength. We have installed a commercially-available Single-photon avalanche diode (SPAD) array on the focal plane of the REVOLT AO testbench at the Dominion Astrophysical Observatory, allowing us to observe the visible PSF of the system at rates up to 96 kHz. This provides a time-resolved view of the PSF at $\sim$100 times the frequency of the AO system itself. We use this detector to analyze the high-frequency behaviour of the PSF of the AO system, including residual tip/tilt and deformable mirror response. We also explore the performance of AO-assisted lucky imaging, in which we average together only the frames which result in the best image quality. We find that high-framerate imaging can significantly improve the PSF beyond the native capabilities of the AO system.
\end{abstract}


\section{INTRODUCTION}
\label{sec:intro}

Modern astronomy relies on the use of highly sensitive adaptive optics (AO) techniques to probe the high--spatial resolution and high-contrast domains from the ground. This requires detectors capable of sensing the wavefront of incoming light on the order of 1 kHz and actively controlling deformable mirrors to flatten the wavefront at the same speed. Yet still, the corrected light has residual distortions evolving both spatially and temporally. To identify how the AO correction can be improved and to fully understand the dynamics of our system, it is thus beneficial to investigate detectors capable of operating at or above the speed of an AO system with near perfect sensitivity and zero noise.

Single-photon avalanche diodes (SPADs) are a promising technology offering high-speed noiseless readout. SPADs are photodiodes reverse-biased above their breakdown voltage, such that an incoming photon will generate a macroscopic current in the diode which is counted digitally. After a photon detection event, the bias voltage is briefly lowered to stop the current, and the bias is increased again so that the diode is ready to detect another photon.\cite{Carrier2023}.

In contrast to conventional photodiodes used in astronomy such as charge-coupled devices, SPADs do not require the total charge collected in potential wells to be converted into digital counts all at once, as each photon event is counted individually with a binary counter. This eliminates the read noise associated with the conversion process, resulting in a nearly photon-counting detector. Photon detections are limited only by the ``dead time'' of the device when the bias voltage is reset, which is around 100~ns per detected photon.

In this Proceeding, we explore the uses of SPAD arrays in AO-assisted observations. In Section \ref{sec:observations}, we describe our experimental setup and observations. Section \ref{sec:psf} shows how the PSF of our AO system changes on fast timescales. Section \ref{sec:luckier-imaging} demonstrates techniques enabled by fast, low-noise detectors that improve the image quality beyond that of AO alone. We discuss additional research directions and state our conclusions in Section~\ref{sec:conclusion}.

\section{OBSERVATIONS}
\label{sec:observations}

In order to investigate the capabilities of high-framerate SPAD observations on-sky, we installed a commercially available SPAD array from Micro Photon Devices called ``Hermes''\footnote{\url{https://www.micro-photon-devices.com/products/arrays/hermes}} on the REVOLT AO testbench at the Dominion Astrophysical Observatory in Victoria, Canada. Hermes is a $32\times 64$ pixel, 150~\um{}--pitch array of SPADs sensitive from $\sim$300--600~nm with a peak photon detection efficiency of 42\%. The fill factor of each pixel is increased from 3\% to 70\% with a built-in microlens array, and each pixel contains an 8-bit digital counter. Hermes is capable of reading a full frame once every 10.4~\us{} (96,000 frames per second), and longer exposures are achieved by digitally stacking individual 10.4~\us{} exposures on-board the device. Because of the noiseless nature of the detector, this digital stacking comes at no extra noise cost relative to long exposures. We find that this device has a median dark current of 103 e$^-$/s/px, corresponding to one dark count per pixel every 930 frames.\cite{Grosson2026}

\begin{figure}[h]
    \centering
    \includegraphics[width=0.9\linewidth]{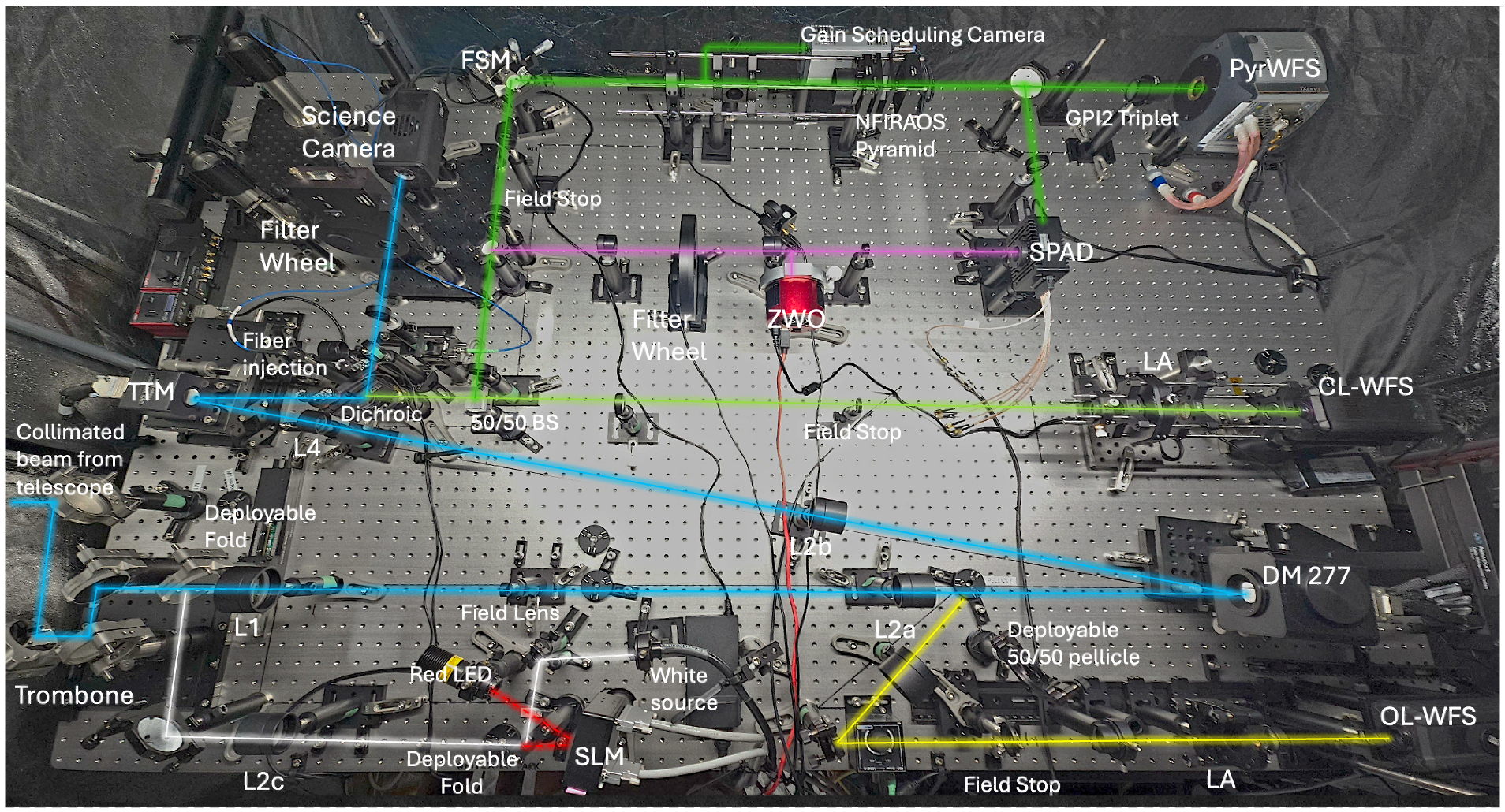}
    \caption{Annotated diagram of the REVOLT AO testbench as of 2026. See Ref. \cite{Jackson2026} for a detailed description of REVOLT.}
    \label{fig:revolt}
\end{figure}

REVOLT\cite{Jackson2024} is a platform for a variety of AO experiments, including a science arm with both a fiber injection unit and a near-infrared camera, a Shack-Hartmann Wavefront Sensor (SH-WFS), and a Pyramid Wavefront Sensor (PWFS). We placed the SPAD in the focal plane of the PWFS arm so that AO correction could be applied using the SH-WFS during observations. With this setup, shown in Figure \ref{fig:revolt}, our detector has a pixel scale of 140 mas/px, with a $4.5 \times 9.0$" field of view. A filter wheel with standard Optolong LRGB filters was included in front of the SPAD to compare performance across wavelengths.

We performed closed-loop observations of Arcturus at 96,000 frames per second while operating the AO system at 1 kHz using the SH-WFS. We obtained approximately 10 seconds of data (1,000,000 frames) in each of the four filters, as well as corresponding dark frames. For each filter, the median dark image was subtracted from each individual frame. A sample 10.4~\us{} exposure is shown in Figure \ref{fig:arcturus-10.4}.

\section{TIME-RESOLVED PSF BEHAVIOUR}
\label{sec:psf}

\begin{figure}[ht]
    \centering
    \href{https://dx.doi.org/10.1117/12.3103095.1}{\includegraphics[width=0.8\linewidth]{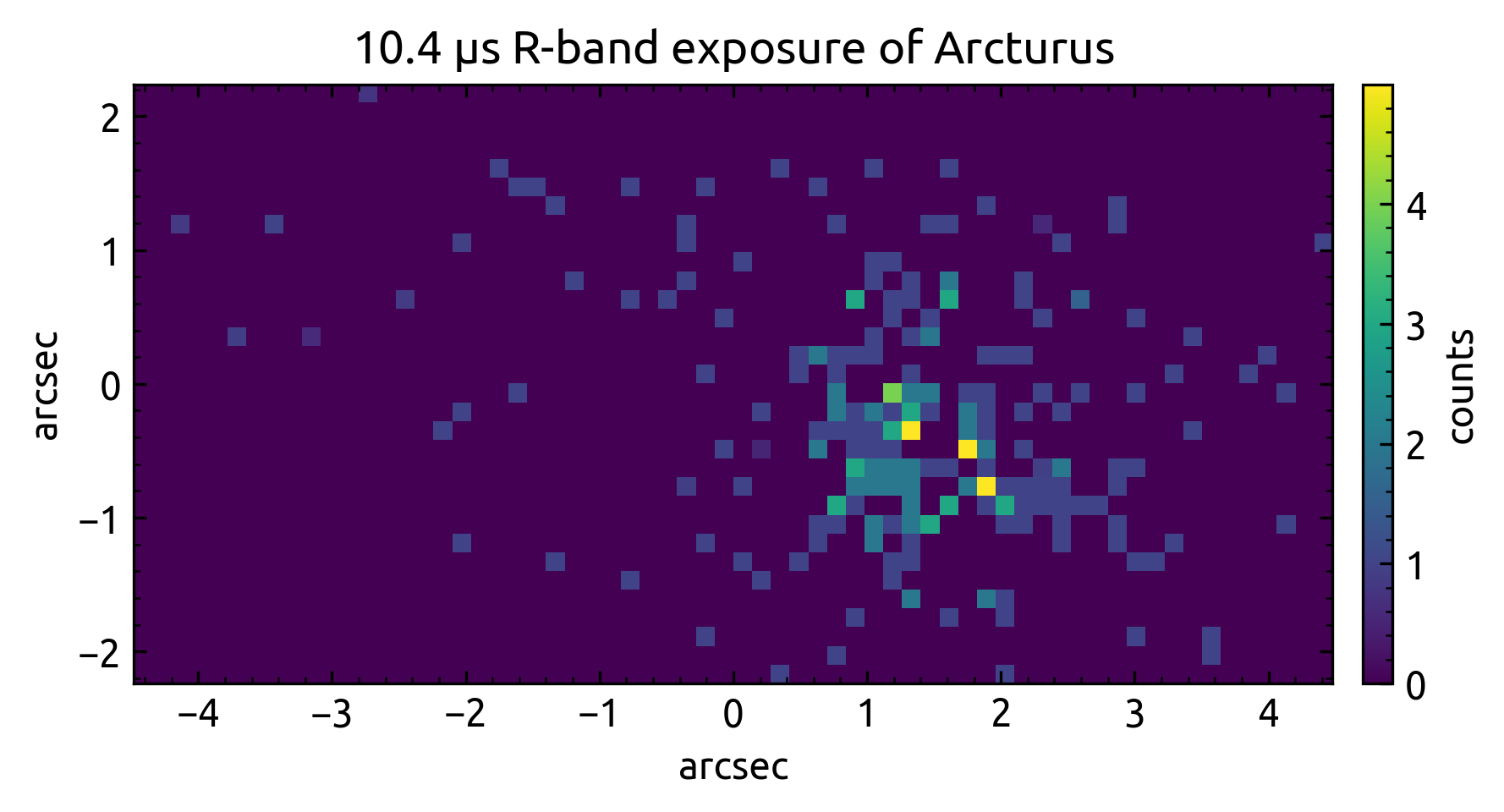}}
    \caption{Video 1. Dark-subtracted 96,000 fps R-band animation of Arcturus using the SPAD on REVOLT. \url{https://dx.doi.org/10.1117/12.3103095.1}}
    \label{fig:arcturus-10.4}
\end{figure}

\begin{figure}
    \centering
    \href{https://dx.doi.org/10.1117/12.3103095.2}{\includegraphics[width=0.8\linewidth]{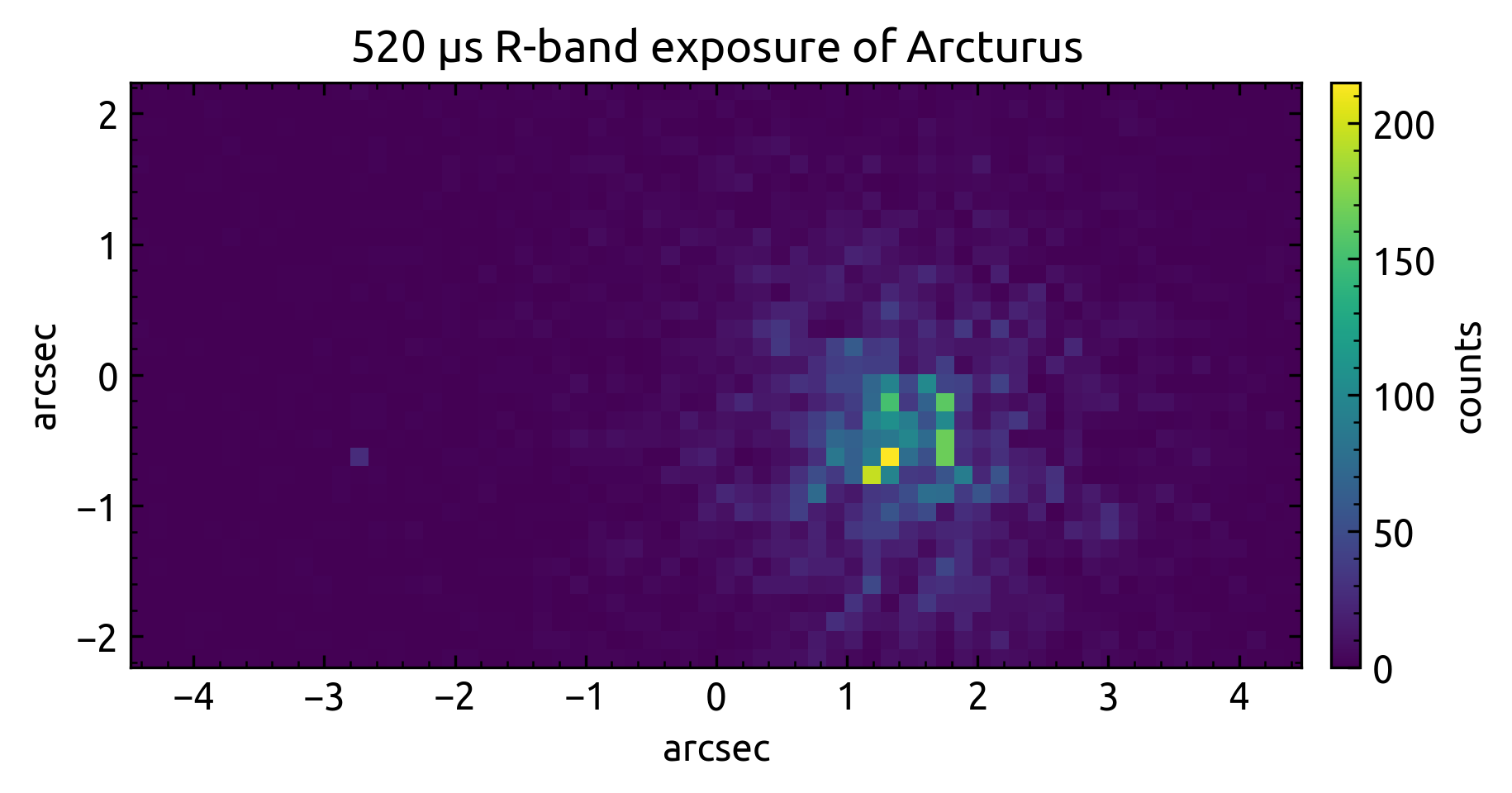}}
    \caption{Video 2. Dark-subtracted 1,900 fps R-band animation of Arcturus, created by summing consecutive 10.4~\us{} frames in groups of 50, resulting in 520~\us{}--equivalent exposures. \url{https://dx.doi.org/10.1117/12.3103095.2}}
    \label{fig:arcturus-520}
\end{figure}

Imaging at such high framerates allows us to observe how the PSF behaves on very short timescales. For example, the video linked in Figure \ref{fig:arcturus-10.4} demonstrates where individual photons land in the PSF, with each pixel only capturing a maximum of around 5 photons per 10.4~\us{} frame. An animation of the PSF in bins of 520~\us{}, shown in Figure \ref{fig:arcturus-520}, shows the residual tip/tilt of the AO system, as well as how the speckle pattern changes over time. We leave analysis of the speckle pattern to future work, focusing this paper on the motion of the PSF.

We measure the motion of the PSF by tracking the location of its centre across each of the 10.4~\us{} frames. This is done by creating an image of the theoretically ideal PSF for REVOLT from an image of the telescope pupil, creating one PSF for each filter. Each on-sky frame is then cross-correlated with the ideal PSF, and the location of the maximum correlation is taken to be the location of the PSF for that frame. To achieve sub-pixel precision, the maximum correlation location is found by fitting a 2-dimensional parabola to the $3\times 3$ grid of elements surrounding the maximum value of the cross-correlation matrix.

\begin{figure}
    \centering
    \includegraphics[width=0.9\linewidth]{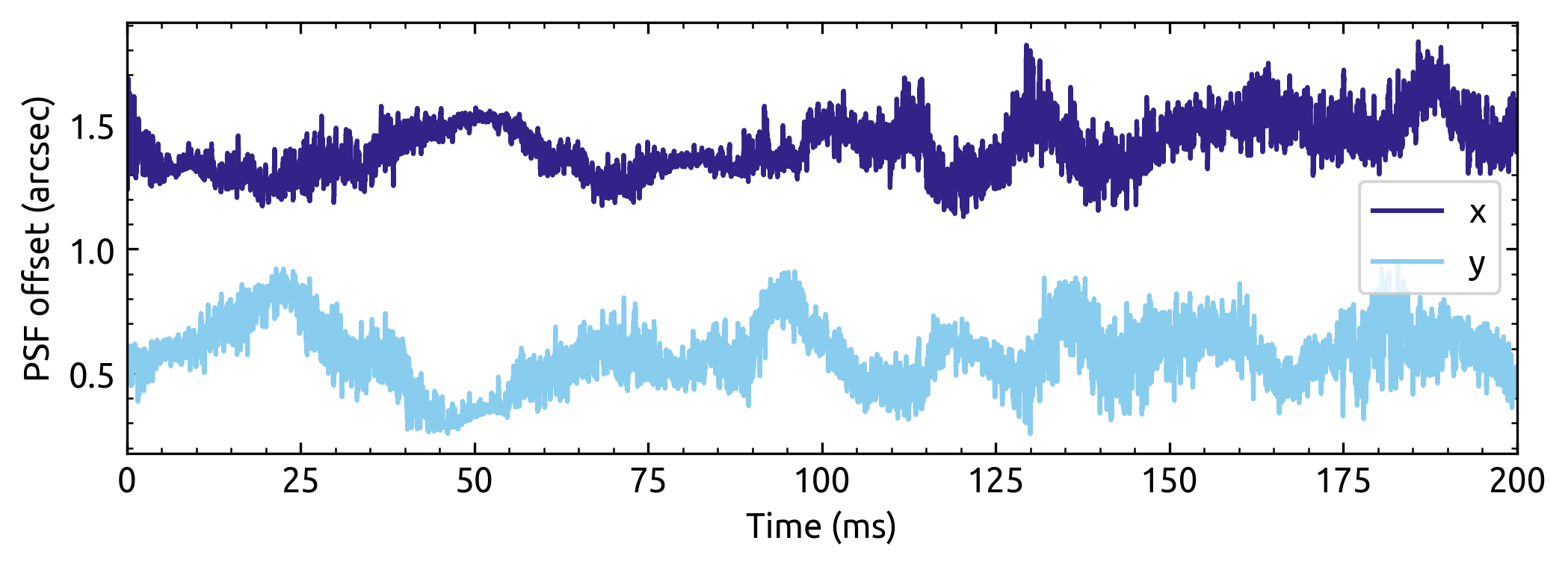}
    \includegraphics[width=0.9\linewidth]{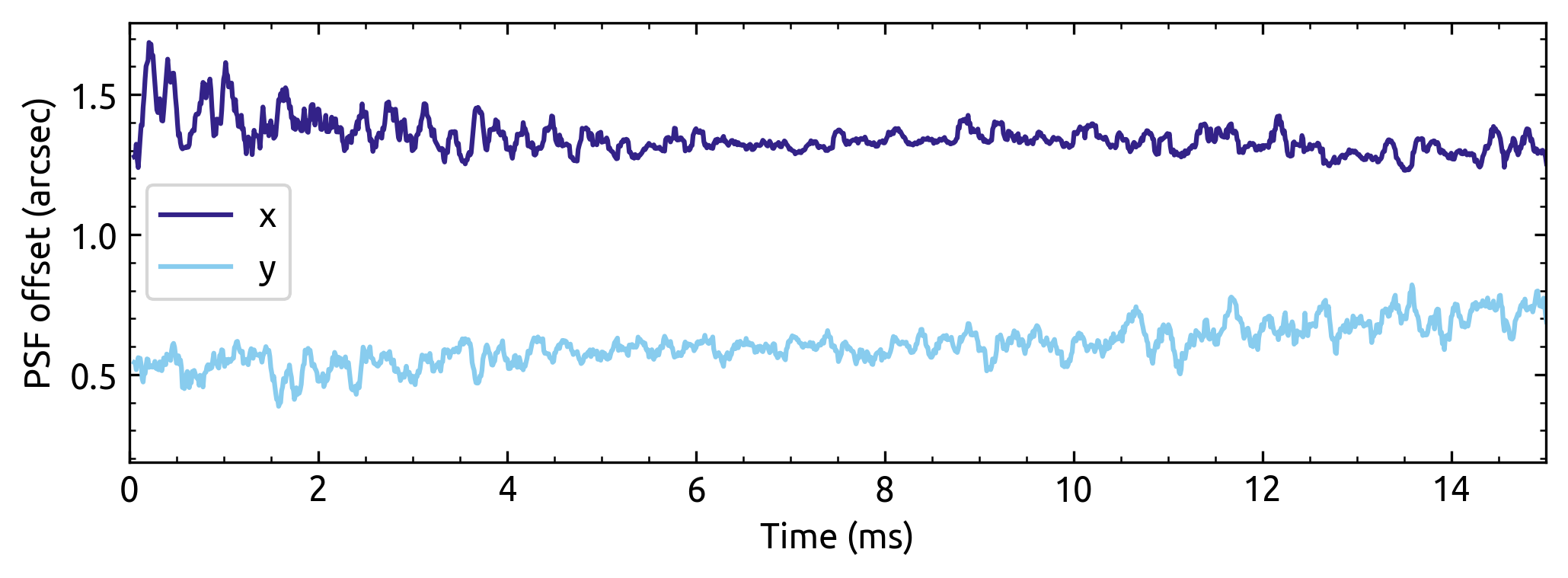}
    \caption{Residual motion of the PSF of Arcturus in the R band after AO correction, for two different timescales.}
    \label{fig:psftrack}
\end{figure}

After boxcar smoothing the time series with an 11-element kernel, we plot the motion of the PSF in the R band in Figure \ref{fig:psftrack}. On longer timescales, there is a residual tip/tilt with variations occurring on the scale of 10~ms, and a RMS variability of 0.8\arcsec{} over the 10~s exposure. On shorter timescales, there are fast oscillations with a principle frequency of $\sim$2000~Hz. Both variations indicate potential avenues for improvement of AO systems, which we explore below.

\subsection{Response of the Deformable Mirror}

\begin{figure}[ht]
    \centering
    \href{https://dx.doi.org/10.1117/12.3103095.3}{\includegraphics[width=0.8\linewidth]{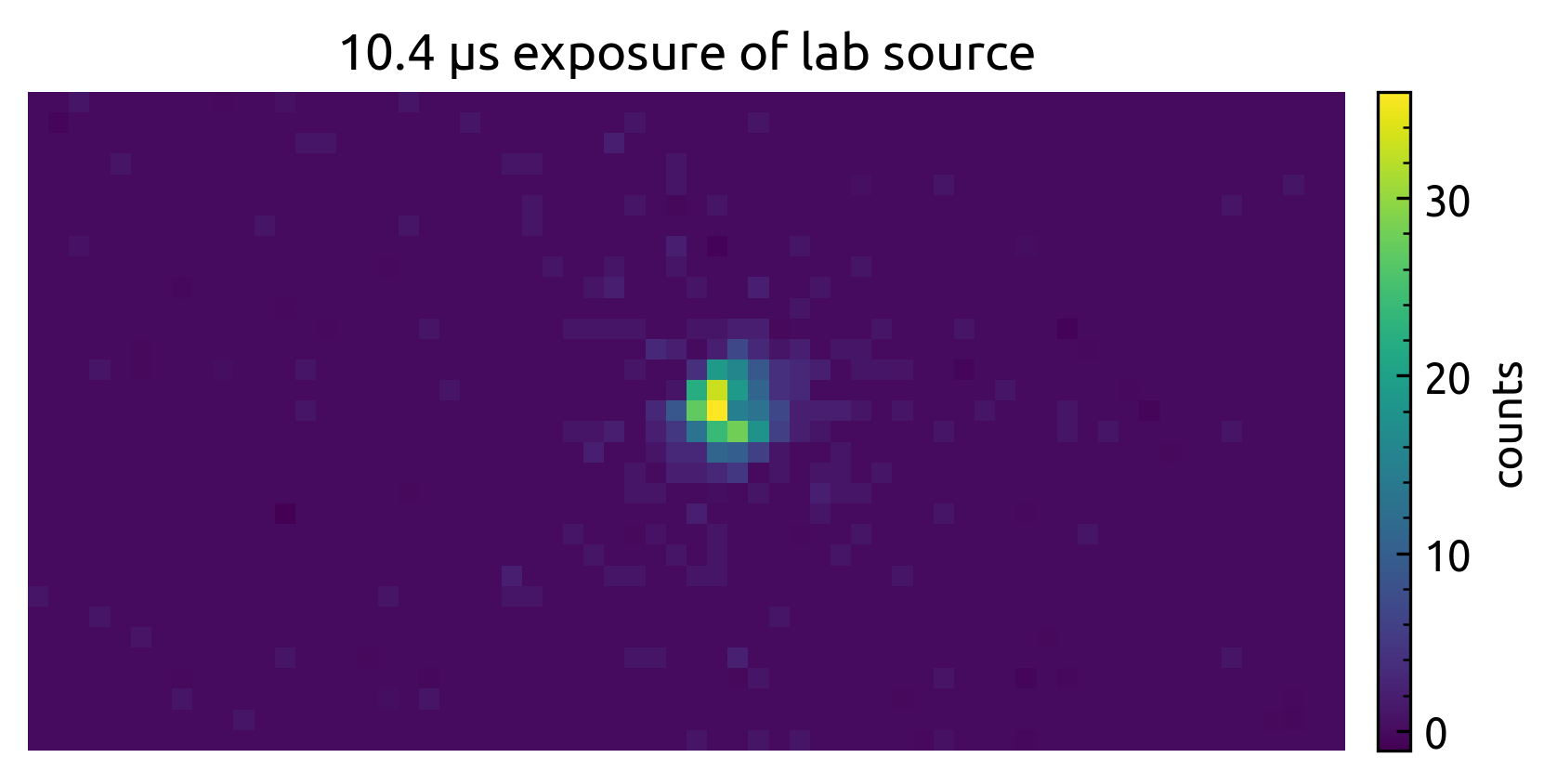}}
    \caption{Video 3. 96,000 fps observation of the white light source as the DM is tipped, animation binned by a factor of 5 to demonstrate the PSF motion. \url{https://dx.doi.org/10.1117/12.3103095.3}}
    \label{fig:lab-10.4}
\end{figure}

A potential limiting factor of any AO system is its deformable mirrors (DMs). The DM provides real-time corrections to the incoming wavefront by moving individual actuators to match its surface to the shape of the measured wavefront, typically at around 1~kHz. In order to identify how the DM itself affects the behaviour of the PSF, we used the internal light source on REVOLT to perform daytime testing. After flattening the DM, we imaged the PSF with the SPAD at 96,000~fps while a tip command was sent to the DM, shifting the PSF by about 6~px on the SPAD. An animation of this is shown in Figure \ref{fig:lab-10.4}, and additional animations at different speeds can be seen at \url{https://github.com/tgrosson/SPAD}.

\begin{figure}[ht]
    \centering
    \includegraphics[width=0.9\linewidth]{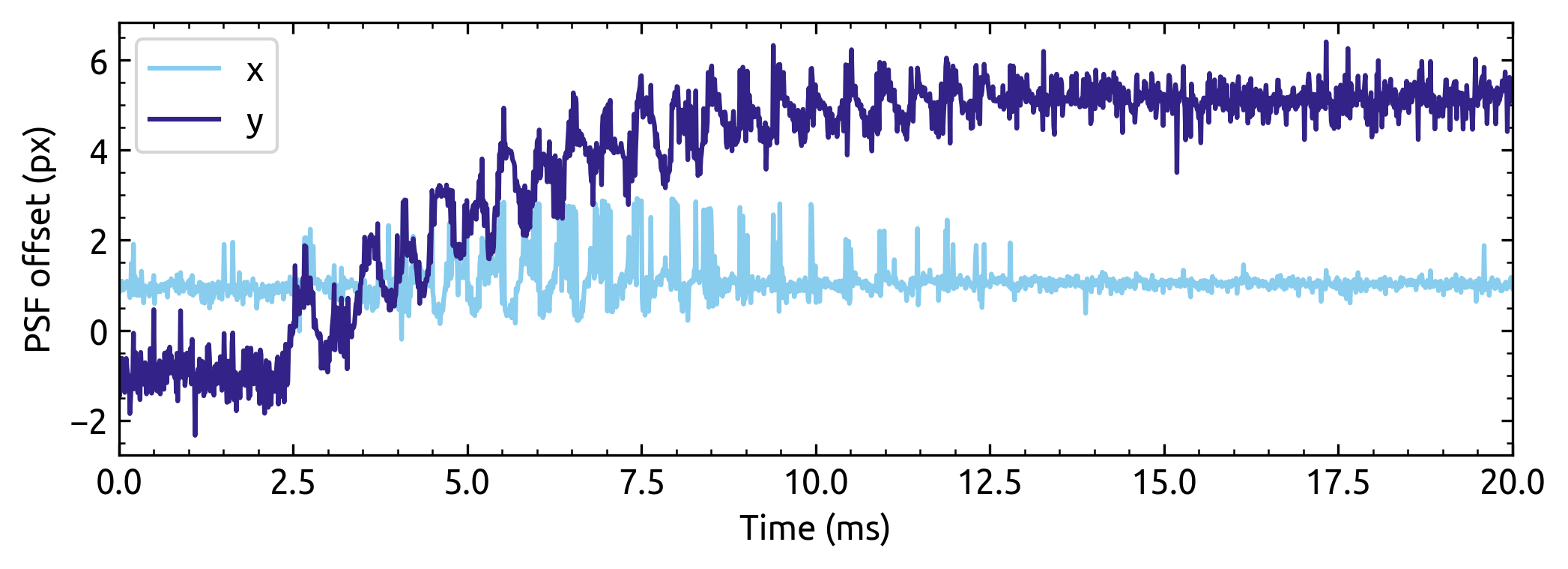}
    \caption{Position of the PSF centre after changing the tip of the DM.}
    \label{fig:dmtrack}
\end{figure}

Although we often consider the response of a DM to be instantaneous, the animations clearly show that this is not the case, as the PSF oscillates toward its new position over several milliseconds. We measure the location of the PSF in each frame using the same method as before, and show the time series in Figure \ref{fig:dmtrack}. This DM, an ALPAO DM277 from 2011, exhibits strong oscillations at around 2000~Hz over the course of its motion, as well as a settling time of around 10~ms.

This motion in response to a single DM command suggests that it may be the behaviour of the DM itself that drives some of the residual motion of the PSF in on-sky observations. A settling time of 10~ms means that the DM may not reach its intended position before needing to move again, resulting in an insufficiently-corrected tip/tilt. This can be addressed to some extent by a dedicated tip/tilt mirror which corrects only the first order wavefront shape, while leaving higher orders to a secondary mirror. A tip/tilt mirror consists of a more rigid body resulting in improved response, though it may still have some oscillations and settling time which result in a residual tip/tilt.

The high-frequency oscillations are intrinsic to the structure of the DM, so they may be more difficult to eliminate. It may however be possible to measure the resonant frequencies of a given DM, and apply DM commands which account for these frequencies. The development of such a technique is a potential area of further study for high-framerate imaging.

\section{IMPROVING THE PSF BEYOND AO}
\label{sec:luckier-imaging}

\subsection{Shift-and-Stack}

\begin{figure}[ht]
    \centering
    \includegraphics[width=0.49\linewidth]{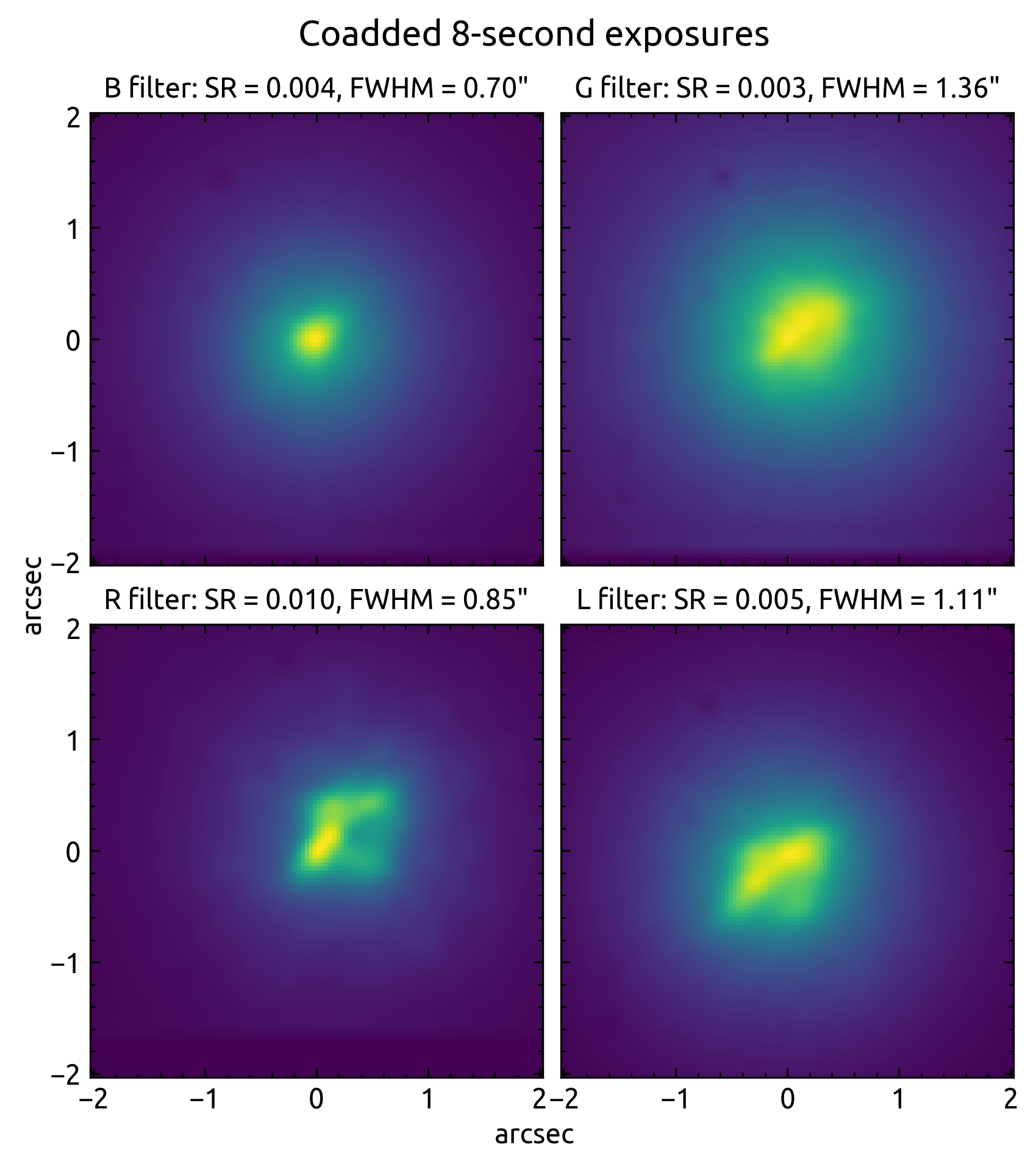}\includegraphics[width=0.49\linewidth]{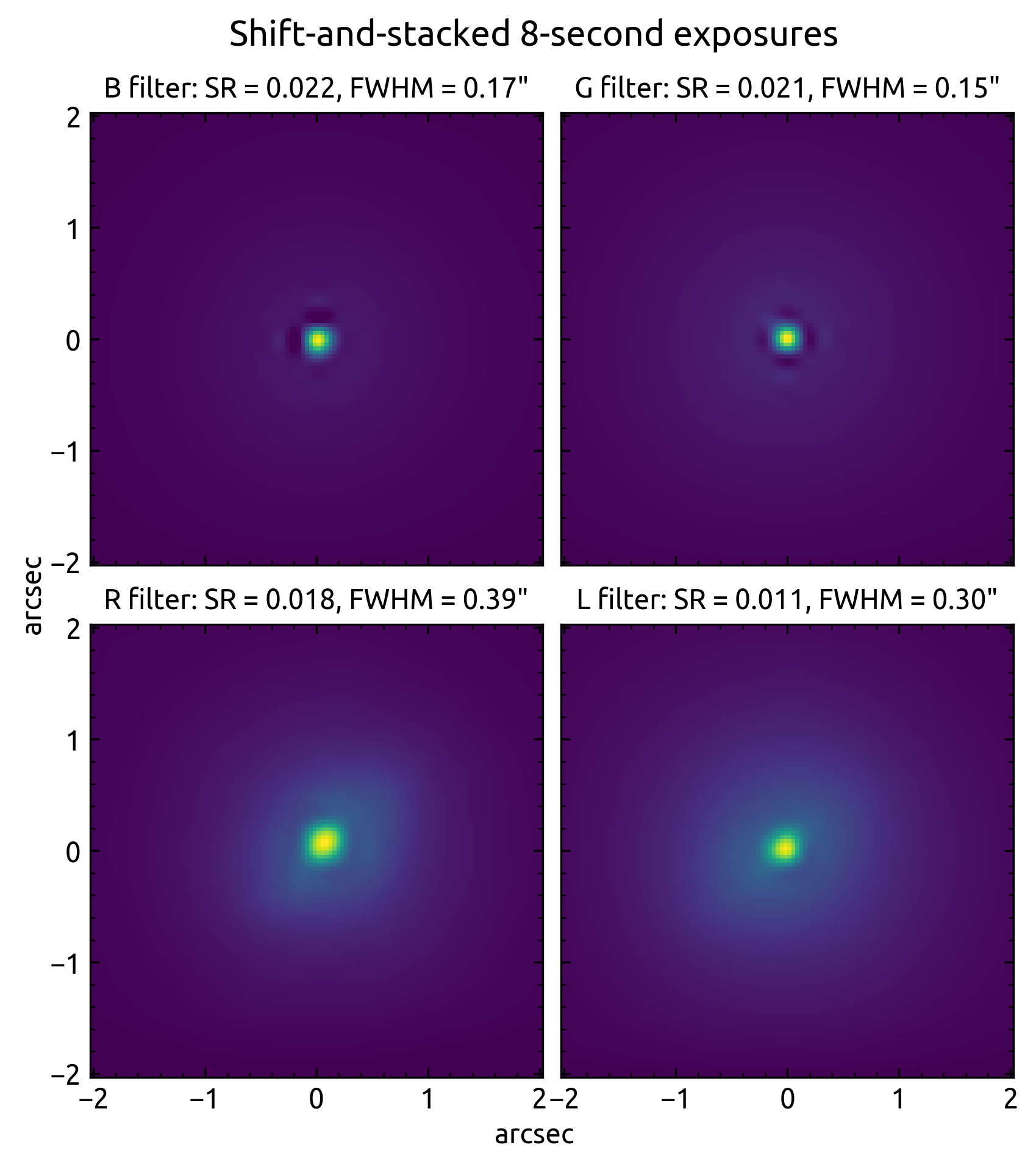}
    \caption{Coadded PSFs of an 8-second exposure of Arcturus in B, G, R, and L filters without (left) and (with) shift-and-stacking each of the 10.4~\us{} frames. For each PSF, the Strehl ratio (SR) and full-width-at-half-maximum (FWHM) are displayed. The asymmetric PSFs on the left may be the result of imperfect telescope tracking and chromatic smearing. Shift-and-stack greatly reduces the effect of the residual tip/tilt.}
    \label{fig:psfs}
\end{figure}

In addition to diagnosing residual tip/tilt in the AO system, high-framerate imaging allows us to apply a post-processing correction to the observations, improving the PSF beyond the native capbilities of the AO system. This is achieved by shifting each frame such that the centre of its PSF, as measured previously, is aligned with all other frames. The resulting coadded image has a significantly more concentrated PSF, as seen in Figure \ref{fig:psfs}. This figure shows the resulting image of an 8--total second exposure without and with the shift-and-stack algorithm for each of the four filters, upsampled by a factor of four to better visualize detail. For each PSF, we show the Strehl ratio (SR)---the ratio of the peak value of the PSF to the peak value of the theoretical PSF---and the full-width-at-half-maximum (FWHM). On our 1.2~m telescope, the perfect, diffraction-limited PSF would have SR~=~1 and FWHM = 77, 92, 109, and 92~mas for the B, G, R, and L filters, respectively.

\begin{figure}[ht]
    \centering
    \includegraphics[width=0.7\linewidth]{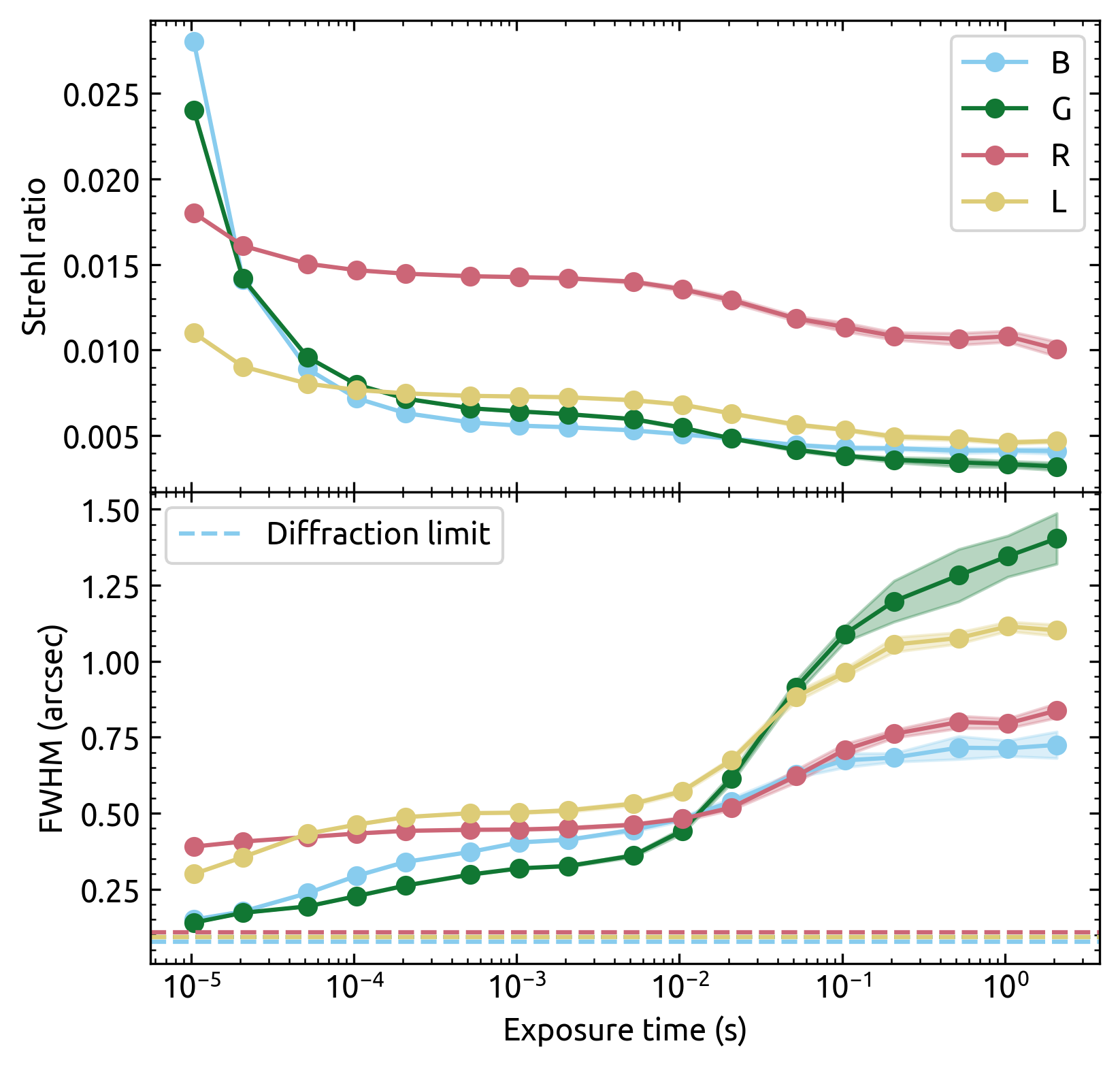}
    \caption{SR and FWHM of an 8-second exposure in each of the four filters as a function of exposure time of each individual frame. Error bars are calculated by bootstrap resampling from the binned frames at each exposure length, and taking the standard deviation of the resulting calculated SR or FWHM.}
    \label{fig:shift-and-stack}
\end{figure}

Although we obtained our data at the maximum detector speed of 96,000~fps, we also investigate how shift-and-stack performs at slower speeds by performing shift-and-stack on sequences of frames which have been binned into longer exposures of various lengths. Because of the noiseless nature of SPAD readout, this is identical to operating the detector at slower speeds. The SR and FWHM in each filter as a function of individual exposure time is shown in Figure \ref{fig:shift-and-stack}.

As seen, it is not necessary to operate a detector at 100 kHz to achieve improvements with the shift-and-stack algorithm. PSF FWHM decreases sharply with exposure times shorter than 0.1~s, with more subtle improvements for exposure times shorter than 10~ms. SR shows a similar improvement for exposure times shorter than 10~ms, although it increases quickly as the exposure time approaches 10~\us{}, reaching \textgreater ~2\% in the B and G filters.

\subsection{Luckier Imaging}

An additional post-processing strategy that can be applied is lucky imaging. Lucky imaging refers to the idea that distortions of a PSF caused by the atmosphere are highly variable, so there are periods of time when these distortions are, by chance, at a minimum. With a sufficiently fast framerate, we can select and coadd only the ``luckiest'' images---those with the least distortion---to achieve a better resolution than the natural atmospheric seeing. Even without the assistance of AO, lucky imaging has been shown to decrease the FWHM of the PSF to near the diffraction limit.\cite{Law2006}

By combining lucky imaging with the REVOLT AO system, we aim to achieve ``luckier'' imaging with higher angular resolution than either lucky imaging or AO-assisted shift-and-stack on their own. After shifting each of the frames to align them, we define the ``luckiest'' frames to be those with the highest peak cross-correlation coefficient values. In each filter, we then coadd the best 50\%, 10\%, and 1\% of frames for each of the previously defined exposure times. As an example, SRs and FWHMs for the G band are shown in Figure \ref{fig:luckier-imaging}, and all four filters are shown in Appendix \ref{sec:appendix}.

\begin{figure}
    \centering
    \includegraphics[width=0.7\linewidth]{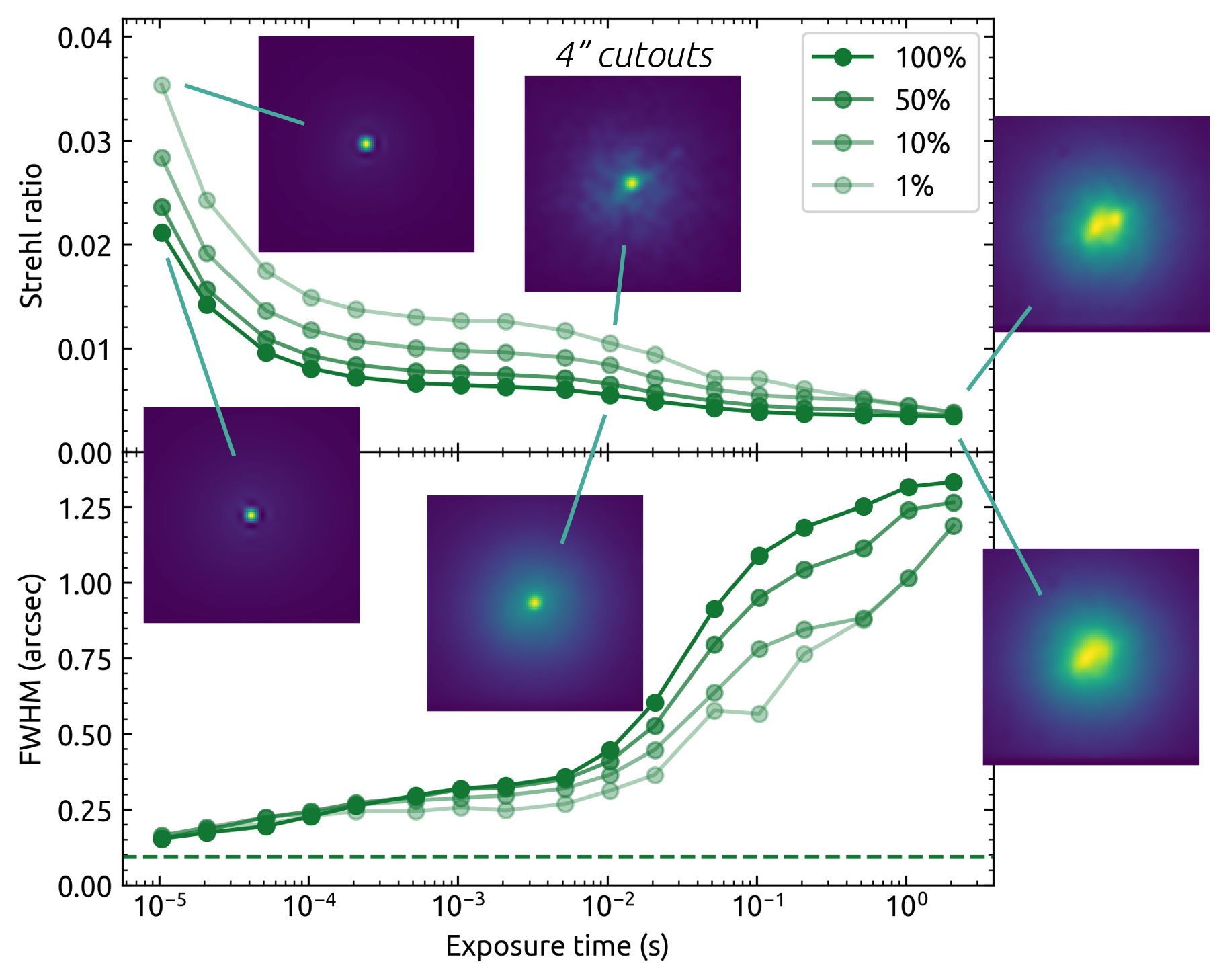}
    \caption{SR and FWHM of an 8-second G-band exposure as a function of exposure time, for different fractions of images kept. ``100\%'' is equivalent to the shift-and-stack only result from above. Sample PSFs for different exposure times and kept fractions are also shown.}
    \label{fig:luckier-imaging}
\end{figure}

Using lucky imaging techniques on our data improves the PSF even further than what shift-and-stack achieved, with SR reaching 4\% in the B-band when using the best 1\% of frames and 10.4~\us{} exposures, a factor of 10 higher than a coadded image without any post-processing. These results greatly exceed the expected performance of REVOLT in the B-band based on the wavefront error budget---we expect $\rm{SR} \ll 1$.\cite{vanKooten2026}

We do not explicitly compare ``luckier'' imaging to lucky imaging without AO due to the seeing at our location (2.1\arcsec{}, $r_0=4.7$~cm) being comparable to the FOV of the detector. However, we note that we are able to achieve FWHMs a factor of 8 smaller than the AO-corrected PSF, in comparison to the study of Ref. \citenum{Law2006}, which achieves a $\sim$3$\times$ improvement over the atmospheric seeing. Ref. \citenum{Law2006} performed their observations with $<1$\arcsec{} seeing in the I-band only, so further work will be necessary to determine the effects of atmospheric seeing over a wider range of wavelengths. We also note that REVOLT throughput and non-common path aberrations are not optimized for visible wavelengths, so we believe even greater results could be achieved with appropriate modifications to the system.

\section{FUTURE WORK AND CONCLUSIONS}
\label{sec:conclusion}

Future work will involve investigation of additional applications of high-framerate imaging with this detector. For example, this detector could be useful in the implementation of time-resolved wavefront sensing (TRWFS) on pyramid wavefront sensors (PWFS). PWFS are more sensitive than traditional Shack-Hartmann wavefront sensors, but the PSF must be modulated around the tip of the pyramid on the order of 1 kHz to improve their linearity, which has the effect of contributing additional photon noise to each PWFS image. TRWFS addresses this problem by imaging faster than the modulation and discarding the images which contribute high amounts of noise and little useful information.\cite{Veran2023} The SPAD detector used in this paper has been used in a preliminary on-sky demonstration of TRWFS, which will be presented in Ref. \citenum{Carrier2026}.

There are also avenues for further research in focal plane imaging with this detector. As discussed in Section \ref{sec:psf}, the motion of deformable mirrors is not instantaneous, instead oscillating for several milliseconds and decreasing the quality of the wavefront correction. Characterization of the resonant modes of deformable mirrors could allow DM commands which account for these resonances.

A detector operating at or above the speed of the AO system could also be used to identify relationships between the PSF and the system operation. REVOLT records telemetry data while operating, including the wavefront sensor gradients and DM commands at full speed with high resolution timestamps. Synchronizing focal plane SPAD observations with the telemetry data may reveal areas in which the AO operation could be improved to result in the best possible PSF.

In this Proceeding, we have demonstrated the on-sky operation of a SPAD detector array on an adaptive optics--enabled telescope. By recording an image every 10.4~\us{}, we are able to understand the behaviour of the visible PSF at speeds faster than the AO system itself. Observations revealed oscillations in the location of the PSF at around 2000~Hz, which can be attributed to resonant behaviour in the DM. Additional residual tip/tilt can be seen varying on timescales around 10~ms, similar to the settling time seen when changing the DM position. These results demonstrate that the behaviour of the DM has a visible influence on the resulting image quality, and correcting for this influence may be necessary for obtaining the best science results possible.

We have also demonstrated that post-processing techniques on high-framerate images can decrease the size of the visible wavelength PSF to near the diffraction limit. Combined with AO, we obtain Strehl ratios $>3\%$ in the B, G, and R filters, and FWHMs around only twice the diffraction limit. Compared to only shift-and-stacking images, lucky imaging improves SR and FWHM by up to a factor of 2. The greatest improvements to FWHM occur with exposure times shorter than 10~ms, while SR is increased most when exposure time is decreased below 100~\us{}. These results show that fast, noiseless detectors such as SPADs have the potential to significantly improve astronomical imaging by pushing angular resolutions to even smaller scales.

\acknowledgments 

REVOLT has been funded by the Astronomy Technology Directorate of the National Research Council of Canada's Herzberg Astronomy and Astrophysics Research Centre.

\bibliography{main}
\bibliographystyle{spiebib}

\appendix

\section{LUCKIER IMAGING PSF RESULTS}
\label{sec:appendix}

\begin{figure}[ht]
    \centering
    \includegraphics[width=0.7\linewidth]{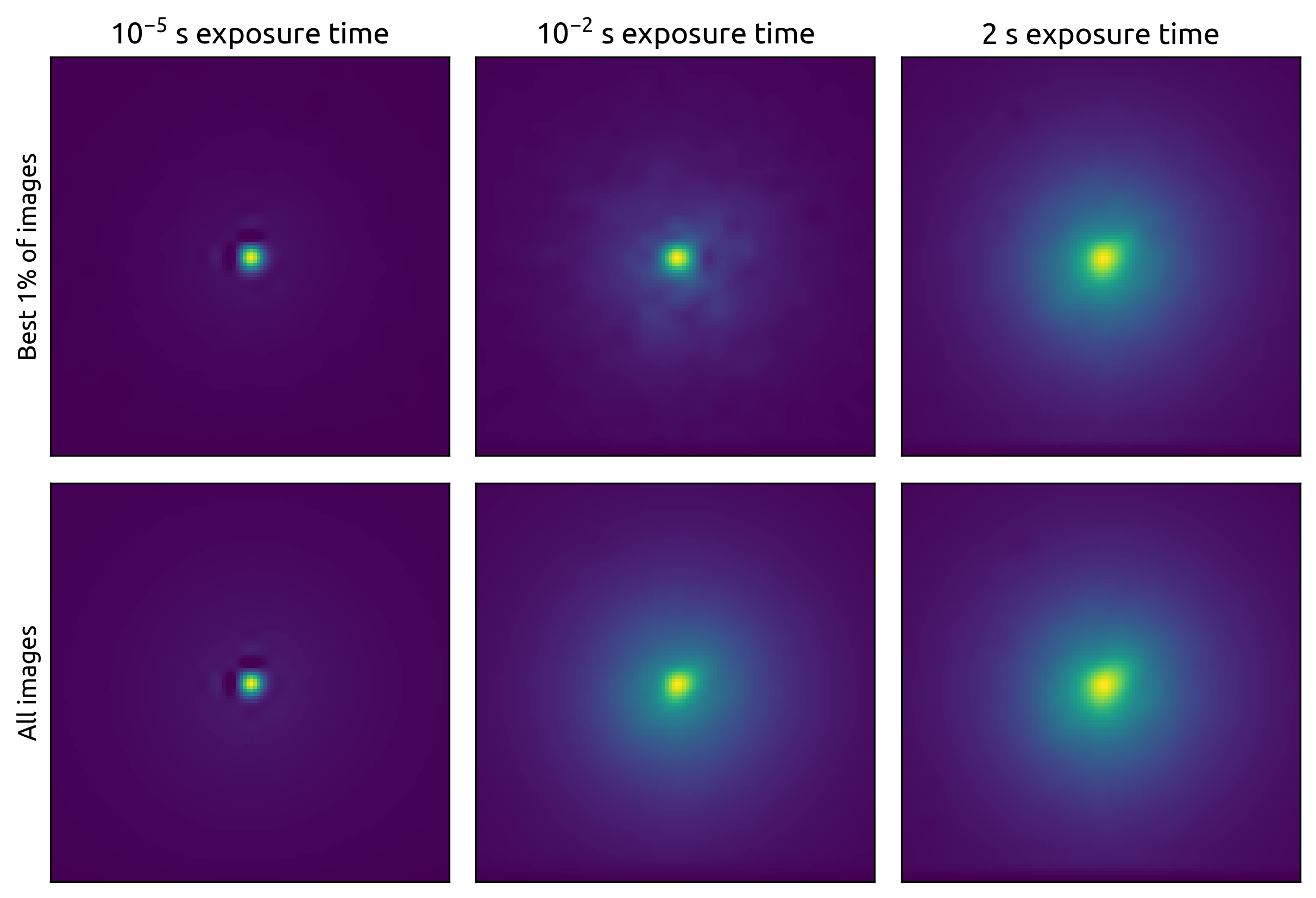}
    \includegraphics[width=0.7\linewidth]{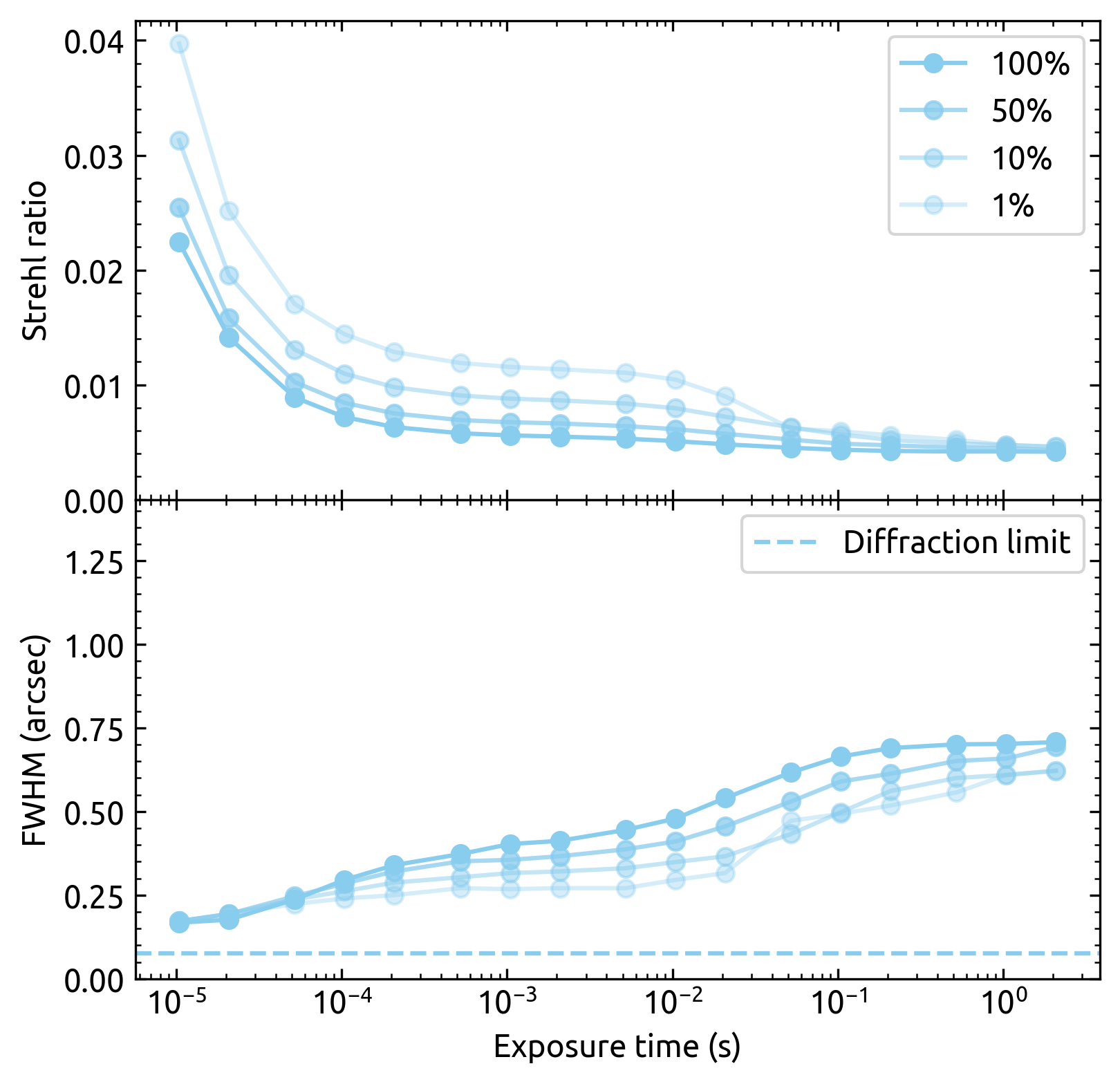}
    \caption{SR and FWHM of an 8-second B-band exposure as a function of exposure time, for different fractions of images kept.}
    \label{fig:luckierB}
\end{figure}

\begin{figure}[ht]
    \centering
    \includegraphics[width=0.7\linewidth]{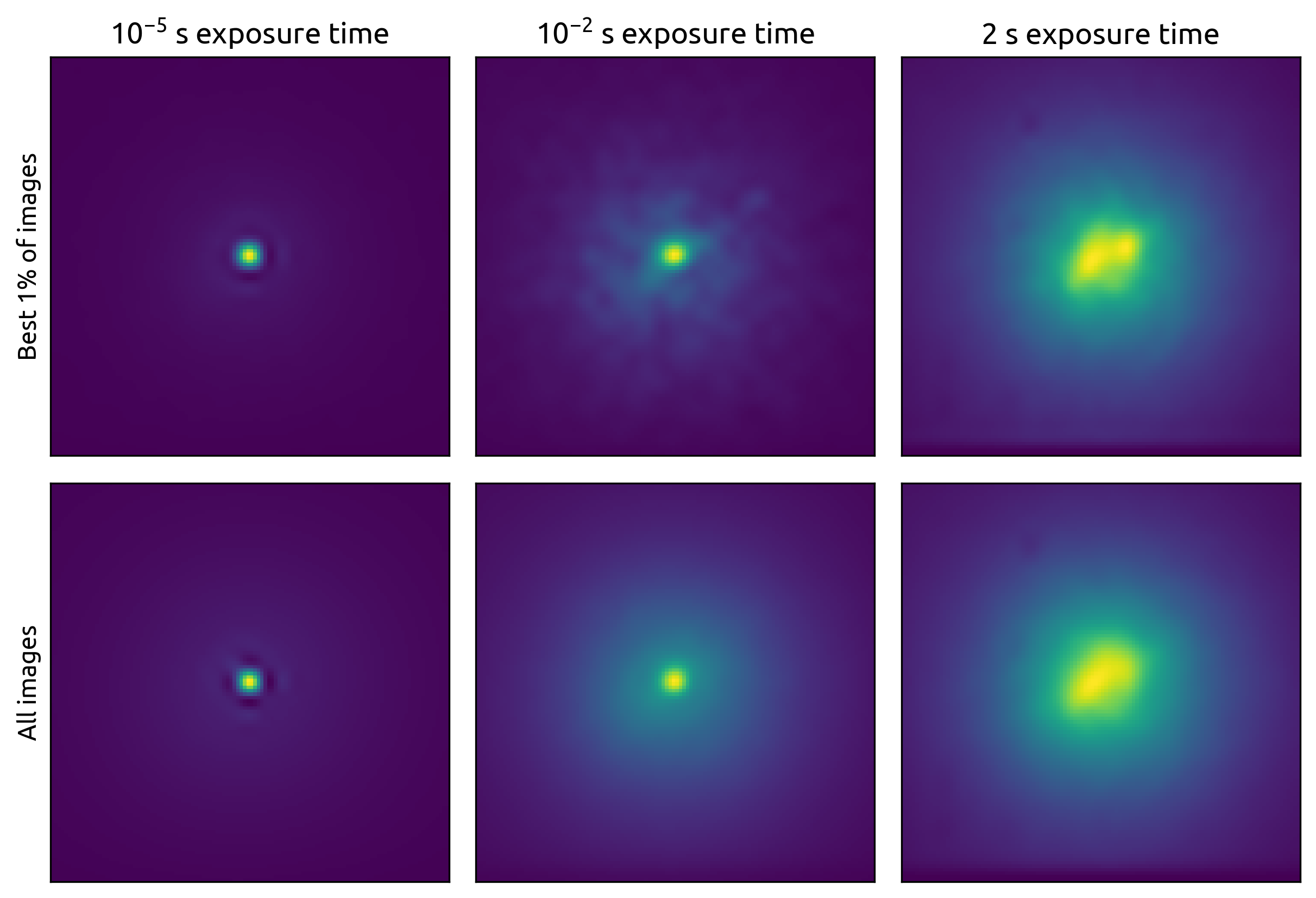}
    \includegraphics[width=0.7\linewidth]{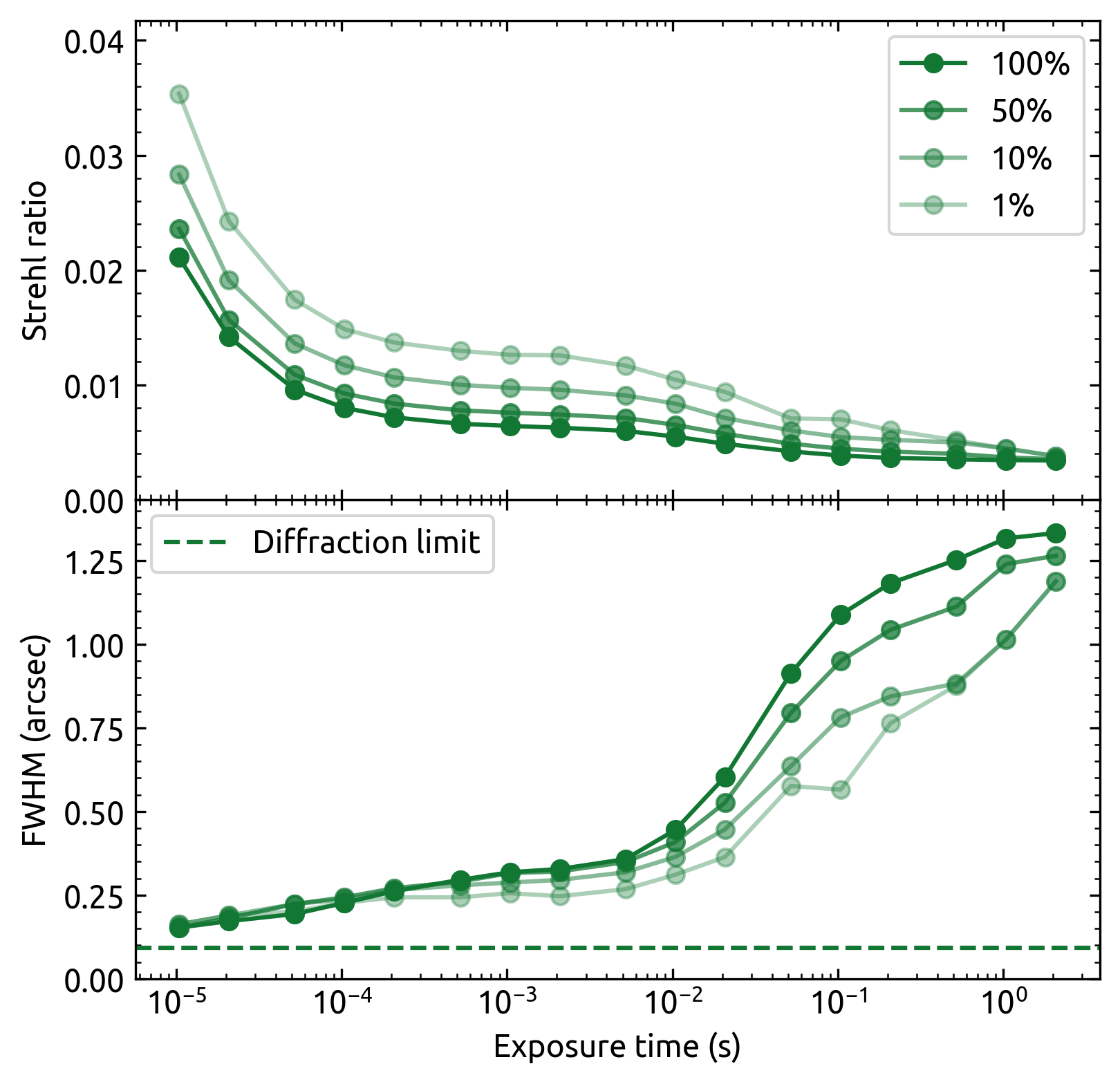}
    \caption{Same as Figure \ref{fig:luckierB}, in the G-band.}
    \label{fig:luckierG}
\end{figure}

\begin{figure}[ht]
    \centering
    \includegraphics[width=0.7\linewidth]{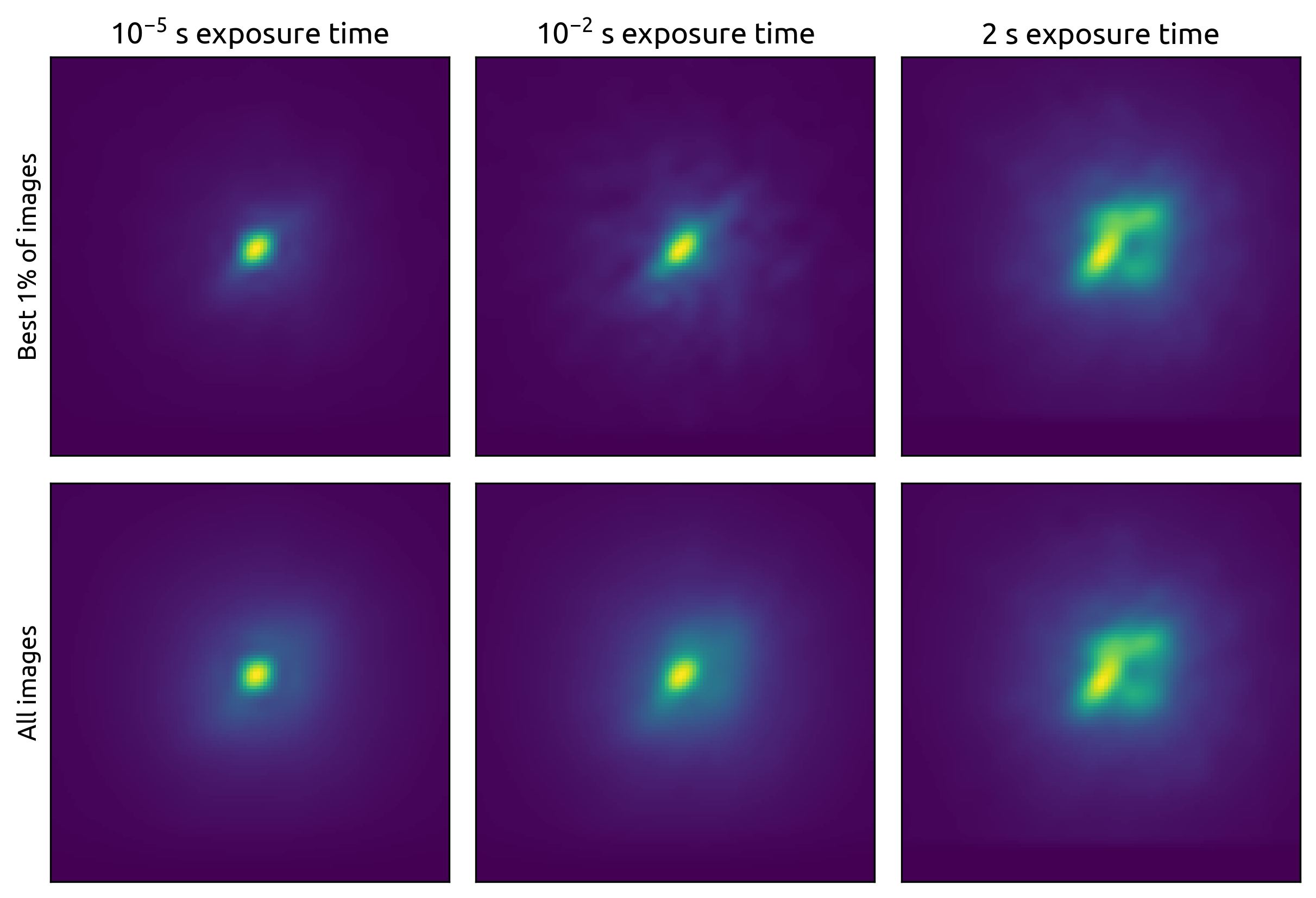}
    \includegraphics[width=0.7\linewidth]{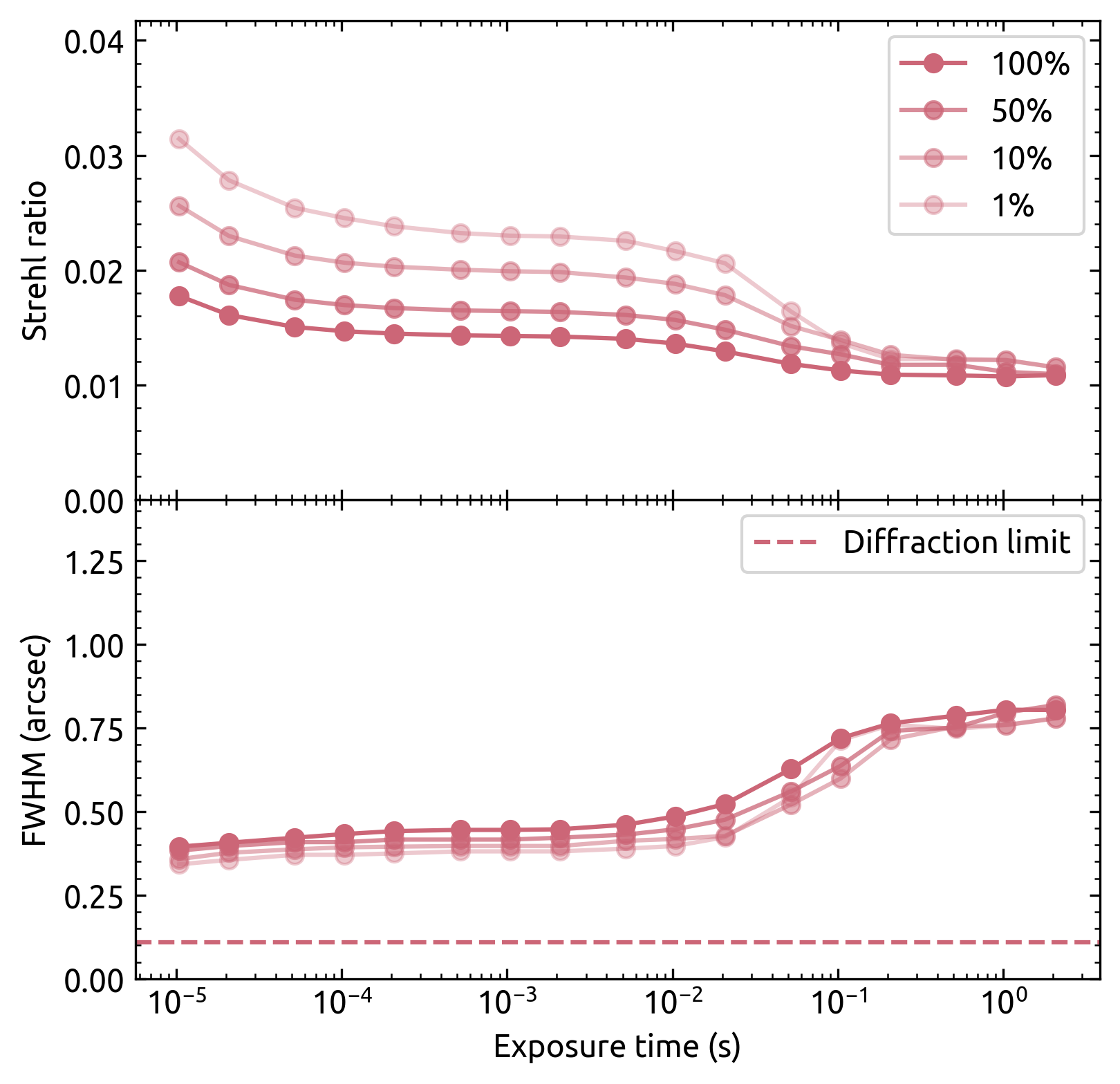}
    \caption{Same as Figure \ref{fig:luckierB}, in the R-band.}
    \label{fig:luckierR}
\end{figure}

\begin{figure}[ht]
    \centering
    \includegraphics[width=0.7\linewidth]{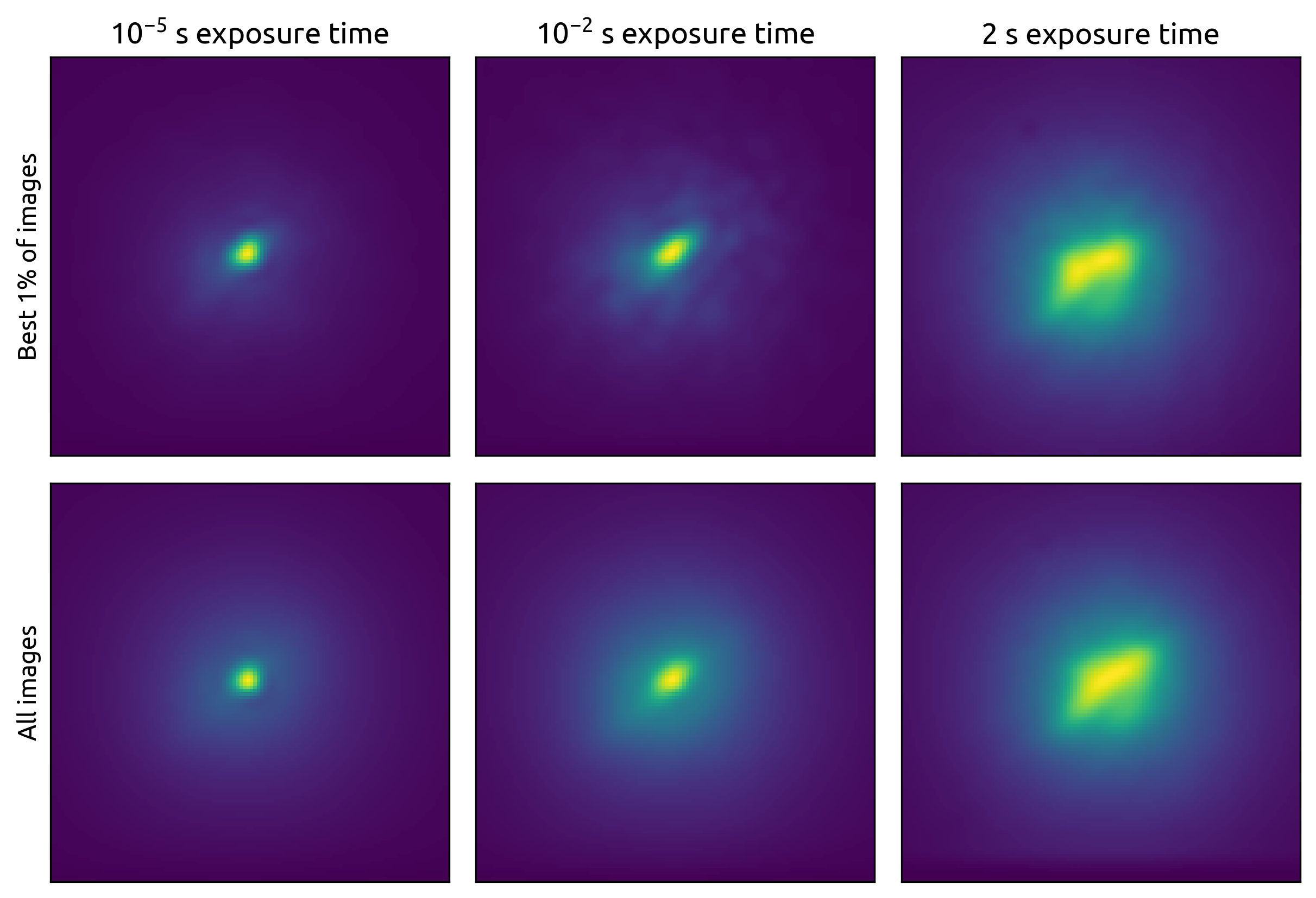}
    \includegraphics[width=0.7\linewidth]{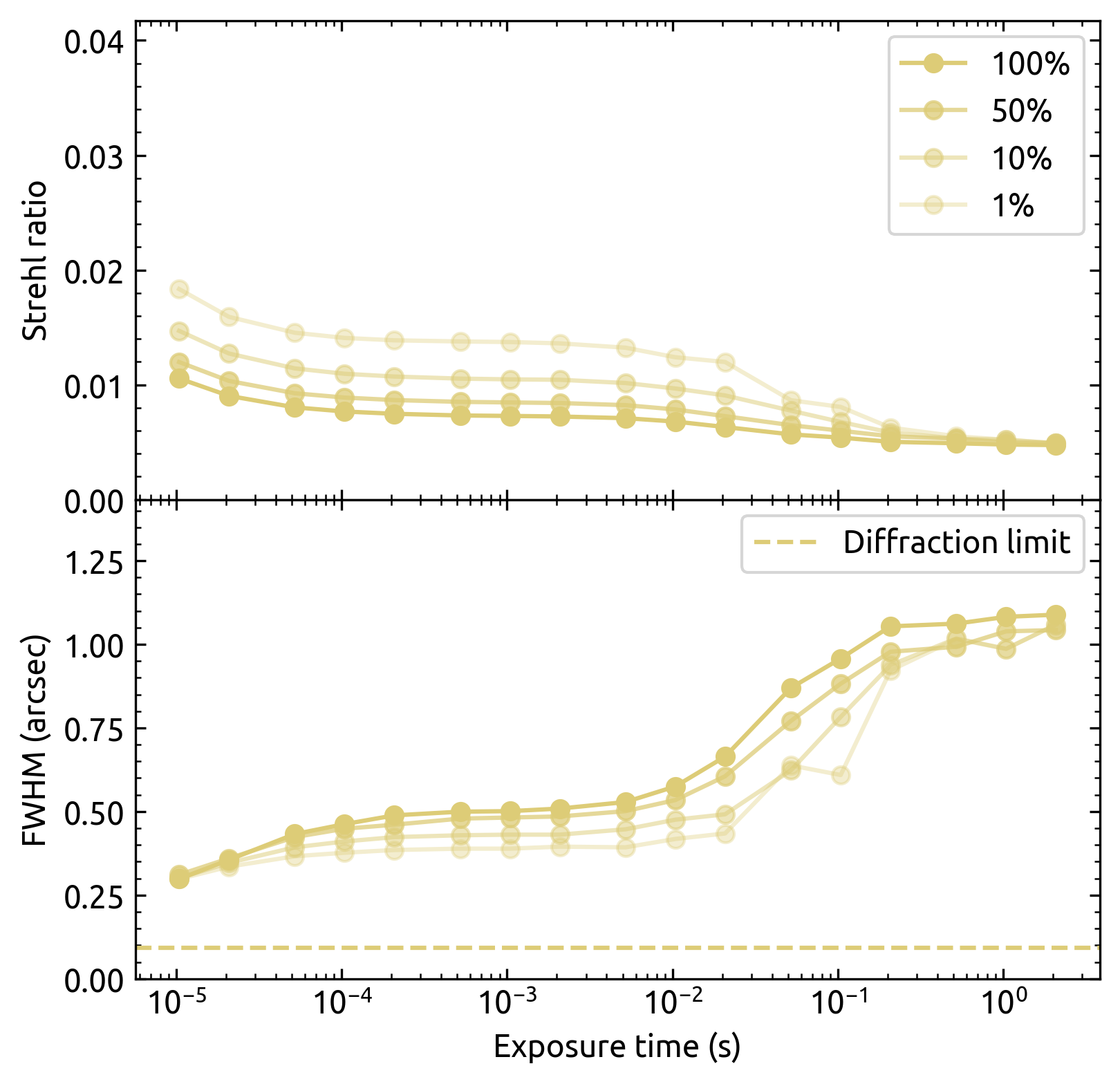}
    \caption{Same as Figure \ref{fig:luckierB}, in the L-band.}
    \label{fig:luckierL}
\end{figure}

Figures \ref{fig:luckierB}--\ref{fig:luckierL} show the SR and FWHM as a function of exposure time and kept image fraction for each of the four filters used. Also shown for each filter are sample PSFs corresponding to different exposure times and kept image fractions.

\end{document}